\RequirePackage{fix-cm}
\documentclass[twocolumn,epjc3]{svjour3}  
\smartqed  
\RequirePackage{graphicx}
\usepackage{csquotes}
\usepackage{amsmath}
\usepackage[british]{babel}

\usepackage{soul}

\usepackage{color}
\usepackage{hyperref}
{
 \definecolor{BLACK}{gray}{0}
 \definecolor{WHITE}{gray}{1}
 \definecolor{RED}{rgb}{1,0,0}
 \definecolor{GREEN}{rgb}{0,1,0}
\definecolor{dgreen}{rgb}{.1,.6,.1}
\definecolor{BLUE}{rgb}{0,0,1}
 \definecolor{CYAN}{cmyk}{1,0,0,0}
 \definecolor{MAGENTA}{cmyk}{0,1,0,0}
 \definecolor{YELLOW}{cmyk}{0,0,1,0}
 \definecolor{aw}{rgb}{0.2,0.5,0.75}
  }
  \hypersetup{
    colorlinks,%
    citecolor=blue,%
    filecolor=blue,%
    linkcolor=magenta,%
    urlcolor=blue
}
\newcommand{\dd}{{\hbox{d}}}
\newcommand{\scal}{{\tiny{\mbox{$(\hspace{-.1em}\phi\hspace{-.1em})$}}}}
\newcommand{\dust}{{\tiny{\mbox{$(\hspace{-.1em}d\hspace{-.1em})$}}}}
\newcommand{\z}{z}
\newcommand{\y}{y}
\newcommand{\x}{x}

\newcommand{\U}{\mathcal{U}}
\newcommand{\W}{\mathcal{W}}
\newcommand{\Vx}{\mathcal{V}_1}
\newcommand{\Vy}{\mathcal{V}_2}
\newcommand{\A}{\mathcal{A}}
\newcommand{\Bx}{\mathcal{B}_1}
\newcommand{\By}{\mathcal{B}_2}
\newcommand{\C}{\mathcal{C}}

\definecolor{MyDarkRed}{rgb}{0.7,0,0}

\journalname{Eur. Phys. J. C}
\begin{document}

\title{
Comment on ``Szekeres universes with homogeneous scalar fields''
}

\titlerunning{Comment on ``Szekeres universes with homogeneous scalar fields''}        

\author{Ismael Delgado Gaspar\thanksref{EmailIsmael,AddrIsmael}
        \and
        Roberto A. Sussman\thanksref{EmailSussman,addrSuss}
        \and 
        David D. McNutt\thanksref{EmailMcNutt,addrMcNutt}
        \and
        Alan A. Coley\thanksref{EmailColey, addrColey} 
}

\thankstext{EmailIsmael}{e-mail: ismidelgado@astro.unam.mx}
\thankstext{EmailSussman}{e-mail: sussman@nucleares.unam.mx}
\thankstext{EmailMcNutt}{e-mail: david.d.mcnutt@uis.no}
\thankstext{EmailColey}{e-mail: aac@mathstat.dal.ca}



\institute{Instituto de Astronom\'\i a, Universidad Nacional Aut\'onoma de M\'exico, AP 70-264, Ciudad de M\'exico, 04510, M\'exico \label{AddrIsmael}
           \and
Instituto de Ciencias Nucleares, Universidad Nacional Aut\'onoma de M\'exico, AP 70-543, Ciudad de M\'exico, 04510, M\'exico \label{addrSuss}
           \and
Faculty of Science and Technology, University of Stavanger, N-4036 Stavanger, Norway 
           \label{addrMcNutt}
           \and
           Department of Mathematics and Statistics, Dalhousie University, Halifax, Nova Scotia, Canada, B3H 3J5  \label{addrColey}
}

\date{Received: date / Accepted: date}

\maketitle


In two recently published articles, Barrow and Paliathanasis (2018, 2019)~\cite{BarrowPaliathanasis,BarrowPaliathanasis2} claim to have found exact solutions of Einstein's field  equations belonging to the class of non-trivial ({\it i.e.}, spatially inhomogeneous) Szekeres models, whose source is a mixture of dust and a homogeneous time-dependent scalar field, where the energy-momentum tensors (EMTs) of both mixture components are independently conserved.  We prove in the present comment that these solutions are inconsistent with the authors' assumptions, as independent conservation of these two mixture components necessarily leads to their solutions belonging to the set of spatially homogeneous subcases of the Szekeres family: Friedmann-Lema\^itre-Robertson-Walker (FLRW) for \\ class I, and Kantowski-Sachs (KS), Bianchi-Behr I or Bianchi-Behr $\mbox{VI}_{\tiny{\mbox{-1}}}$ for class II.  

It is straightforward to understand the motivation behind the papers by Barrow and Paliathanasis: 
they
attempt to generalize the successful $\Lambda$-CDM model by considering the dust mixture component as inhomogeneous cold dark matter (CDM) and the time-dependent scalar field as dynamical dark energy replacing the cosmological constant. However, this generalization is incompatible with independent conservation of the sources, though it is compatible (at least formally) if there is a non-gravitational interaction between them.

It is possible that the authors missed this inconsistency because the Szekeres coordinates they used obscured the fact that their solutions belong to the spatially homogeneous subclass, as these are different coordinates from those conventionally used to describe these spacetimes. However, the Szekeres class of models are characterized in a coordinate independent manner~\cite{kras2} by: \textbf{(i)} A geodesic and irrotational perfect fluid source, \textbf{(ii)} a purely electric Petrov D Weyl tensor and \textbf{(iii)} a shear tensor with two equal eigenvalues whose eigensurface coincides with the eigensurface of the Weyl tensor.  The necessary and sufficient condition for their FLRW limit is the vanishing of the shear tensor, even if this limit is expressed in unconventional coordinates systems that do not explicitly manifest spatial homogeneity and isotropy~\cite{kras2}.
A complete invariant classification can also be obtained using the Cartan-Karlhede algorithm \cite{Kramer}. This has been previously applied to the quasi-spherical Szekeres solutions \cite{CLM2019}.

The standard line element of the Szekeres class of models in comoving coordinates is~\cite{kras2}:
\begin{equation}\label{Eq:GenSzekereLinElem}
\dd s^2=-\dd t^2+e^{2 \alpha(t, x^i)}\dd \z^{2}+e^{2 \beta(t, x^i)}\left(\dd \x^{2}+\dd \y^{2}\right) \ ,
\end{equation}
with $u^\alpha =\delta^\alpha_0$ and $\dot u_\alpha = 0$, implying that $p=p(t)$. We examine class I ($\beta_{,\z}\neq 0$) and class II ($\beta_{,\z}= 0$) separately below. 

For class I, the field equations yield:
\begin{eqnarray}
e^\beta=\Phi e^{\nu (x^i)}  ; \;\;
e^\alpha= \mathcal{F} (\z)\Phi \beta_{,\z} \equiv 
\mathcal{F} \left(\Phi_{,\z} +\Phi \nu_{,\z} \right) , 
\\
e^{-\nu}\!= \A(\z) \left(\x^{2}+\y^{2}\right) +2 \Bx(\z) \x
+2 \By(\z) \y + \C(\z)  ,\label{nudefeq1}
\\
4 \left(\A \C - \Bx^2-\By^2\right)= \left[1/\mathcal{F}(\z)^2+k(\z)\right] , \label{rest1}
\end{eqnarray}
where $k$, $\mathcal{A}$, $\mathcal{B}_1$, $\mathcal{B}_2$, and $\mathcal{C}$  are arbitrary functions, and
the inhomogeneous scale factor $\Phi(t,\z)$ satisfies,
\begin{equation}
2 \ddot{\Phi}/\Phi+(\dot{\Phi}/\Phi)^2 +8 \pi p(t) +k(\z)/\Phi^2=0 \ ,\label{Eq:PhittC-i} 
\end{equation}
while the shear tensor is given by,
\begin{equation}\label{Eq:ShearC-I}
\sigma^{\mu}_{~\nu}=\Sigma\times \mbox{Diag}\left(0,-2,1,1\right) ;
\Sigma =-\frac{1}{3}\frac{\dot{\Phi}_{,\z}-\dot \Phi \Phi_{,\z}/\Phi }{    \Phi_{,\z}+\Phi \nu_{,z} } .
\end{equation}

For class II we have that
\begin{eqnarray}
e^{\beta}&=&a(t) e^\nu, \;\; e^\alpha=a(t) \sigma(x^i)+\lambda(t,\z),\quad
\\
\sigma&=&e^\nu 
\left\lbrack (\U/2) \left(\x^2+\y^2\right)
+\Vx \, \x 
+ \Vy \, \y  + 2 \W\right\rbrack,
\label{SubEq:sigmaCii}
\\
e^{\nu}&=& \left[1+  (k_0 /4)(\x^2+\y^2)\right]^{-1}\ , \; k_0=\mbox{const.},
\label{SubEq:etonuCii}
\end{eqnarray}
where $\U$, $\Vx$, $\Vy$ and $\W$ are arbitrary functions of $\z$, and $a$ and $\lambda$ are solutions of the following equations:
\begin{eqnarray}
2 a \ddot a+\dot a^2+ 8\pi a^2 p+ k_0&=&0,\label{Eq:att_c-ii}
\\
\ddot\lambda a+\dot \lambda \dot a+\lambda \ddot a + 8\pi a\, \lambda  \,p&=&\U+k_0 \W .\label{Eq:lambdatt_c-ii}
\end{eqnarray}
The shear tensor takes the form of~\eqref{Eq:ShearC-I} with
\begin{equation}
\Sigma=(-1/3)(\dot{\lambda}-\lambda \dot{a}/a)/(a\sigma+\lambda) .
\label{Eq:SigmaCii}
\end{equation}
The FLRW limit of the models is recovered when  
\begin{eqnarray}
&&\hbox{Class I:} \;\; \Phi=a(t) f(\z), \; k=k_0 f^2,  \; k_0=\hbox{const.} \;
\label{RWbetaneq0Cond1}
\\
&&\hbox{Class II:} \;\; \lambda=0, \quad \mathcal{U}=-k_0 \W  .
\end{eqnarray}

Consider the fluid mixture of dust and homogeneous scalar field:  
\begin{eqnarray}
\label{Eq:TabSF}
T^{\mu \nu}&=& T^{\mu \nu}_{_\dust}+T^{\mu \nu}_{_\scal},
\quad \hbox{with} \quad T^{\mu \nu}_{\dust} = \varrho_{_\dust}\,u^\mu u^\nu, 
\\
\hbox{and} &{}&  T^{^\scal}_{\mu \nu} = \phi_{,\mu} \phi_\nu - g_{\mu \nu} \left[  \phi_{,\gamma} \,  \phi^{,\gamma}/2 + V(\phi)\right] ,\label{Eq:TabSF2}
\end{eqnarray}
having the form of a perfect fluid  with density and pressure given by 
\begin{eqnarray}\label{Eq:varsSF}
e_{_\scal}(t)= T^{^\scal}_{\mu \nu} u^\mu u^\nu = \dot{\phi}^2 /2+ V(\phi) ,
\\
p_{_\scal}(t)= T^{^\scal}_{\mu \nu} h^{\mu\nu} /3= \dot{\phi}^2/2 - V(\phi)  .
\end{eqnarray}
Demanding the independent conservation of the dust and scalar field components,  
$T^{\mu \nu}_{_\dust ; \nu}=T^{\mu \nu}_{_\scal ; \nu}=0$, leads to the Klein-Gordon equation 
\begin{equation}
\ddot{\phi} + \left(\dot \alpha + 2 \dot \beta \right) \dot{\phi} + \partial_\phi V = 0  ,
\label{Eq:ScalFieldEvol}
\end{equation}
whose consistency implies that the term in parentheses should be a function of time only,
\begin{equation}\label{Eq:ScalFieldConstraint}
\dot \alpha + 2 \dot \beta = \zeta(t) \implies
\alpha + 2  \beta = 3 \ln b(t) + \vartheta(x^i) \ .
\end{equation}
For class I models this condition implies a solution with a separable scale factor, $\Phi=a(t) f(\z)$, which covariantly defines the FLRW limit via the vanishing of the shear tensor. Barrow and Paliathanasis obtained a separable scale factor for class I solutions in~\cite{BarrowPaliathanasis}, but they failed to recognize that the line-element in their equations (16) in~\cite{BarrowPaliathanasis} and (15) in~\cite{BarrowPaliathanasis2} was precisely the FLRW subcase,  claiming instead that it represented an inhomogeneous spacetime as in~\cite{Sze75}.

For class II models, condition~\eqref{Eq:ScalFieldConstraint} splits the solutions into two cases. \textbf{Case (a)}: the FLRW limit of this class. And, \textbf{Case (b)}: $e^\alpha= c(t) \sigma(\z)$, $ e^{\beta}=a(t) e^\nu$, where $\sigma(\z)$ is arbitrary and $c(t)$ obeys,
\begin{equation}
\ddot c/c + \ddot a/a +(\dot a \dot c)/(a c)+8\pi p_{\scal}=0 \ .
\label{Eq:ddc:C-ii}
\end{equation}
Depending on the value of  $k_0$, the spacetime is~\cite{SzaSpatialConfFlat1978} \textbf{ (i)} $k_0=0 $: Bianchi-Behr type I; \textbf{(ii)} $k_0=1 $: KS, or \textbf{ (iii)} $k_0=-1$: Bianchi-Behr type $\mbox{VI}_{\tiny{\mbox{-1}}}$. These results can be summarized as follows:
\begin{quote}
Any Szekeres solution whose source is a mixture of independently conserved inhomogeneous dust and, homogeneous perfect fluid is spatially homogeneous.
\end{quote}
This result was obtained in~\cite{LimaTiomno}, where the authors looked at Szekeres class II models with the EMT \eqref{Eq:TabSF}, but also considered interactions among components with \eqref{Eq:TabSF2} replaced by a perfect fluid with a barotropic equation of state $p(t)=(\gamma-1)e_{_\scal}(t)$  where $\gamma$ is constant.

Interactive mixtures of the type \eqref{Eq:TabSF} with scalar fields interacting with dust were examined in~\cite{sussInt} for the spherically symmetric subcase of class I models. See~\cite{alanInt} for a general reference on the subject. Also, class II models can be generalized to Petrov I solutions admitting a sufficiently general EMT to be compatible with inhomogeneous scalar fields~\cite{NajeraSussman}.
Moreover, the Relativistic Zel'dovich Approximation furnishes a suitable non-linear method to include  scalar fields within the context of Szekeres models~\cite{rza}.



\vspace{-5pt}
\begin{acknowledgements}
The work of IDG is supported by the DGAPA-UNAM postdoctoral grants program; he also acknowledges support from Grants DGAPA-UNAM (IN112019) and SEP-CONACYT 282569 and CB-2014-01 No. 240512.  
AAC is funded by NSERC.
RAS acknowledges support from PAPIIT-DGAPA RR1070 and IA102219.
\end{acknowledgements}
\vspace{-20pt}

\newcommand\eprintarXiv[1]{\href{http://arXiv.org/abs/#1}{arXiv:#1}}

\end{document}